\documentclass{article}

\usepackage{amsmath}
\usepackage{amsfonts}
\usepackage[noend]{algorithmic}
\usepackage{float}
\usepackage{xspace}

\floatstyle{ruled}
\makeatletter
\renewcommand\floatc@ruled[2]{{\@fs@cfont Procedure} #2\par}
\makeatother
\newfloat{algorithm}{tb}{lis}
\floatname{algorithm}{Procedure}

\def\01{\{0,1\}}
\newcommand{\ket}[1]{\ensuremath{|{#1}\rangle}} 
\newcommand{\bigO}{\ensuremath{\operatorname{O}}}

\newcommand{\N}{\ensuremath{\mathbb N}}
\newcommand{\R}{\ensuremath{\mathbb R}}
\newcommand{\expect}{\ensuremath{\operatorname{E}}}

\newcommand{\etal}{et~al.\@\xspace}
\newcommand{\ie}{i.e.,\xspace}
\newcommand{\eg}{e.g.,\xspace}
\newcommand{\lb}[1]{\ensuremath{\| #1 \|}}
\newtheorem{theorem}{Theorem}
\newtheorem{corollary}{Corollary}
\newtheorem{lemma}{Lemma}
\newtheorem{definition}{Definition}
\newenvironment{proof}[1][Proof]{%
  \begin{trivlist}{}{\setlength{\topsep}{0cm}\setlength{\partopsep}{0cm}}
  \item \textbf{#1.\@}\hspace*{1ex}\ignorespaces}%
  {\phantom{.}~\hfill$\Box$\end{trivlist}}
\newcommand{\qed}{}

\newcommand{\robustdeg}{\ensuremath{\operatorname{rdeg}}}
\newcommand{\approxdeg}{\ensuremath{\widetilde{\operatorname{deg}}}}

\newcommand{\prbl}[1]{\textsc{#1}}
\newcommand{\procref}[2][o]{\textup{#2}}
\newcommand{\procdef}[2][o]{\label{proc:#2}}
\newcommand{\ignore}[1]{}

\def \xor {\oplus}

\begin{document}

\title{Robust Polynomials and Quantum Algorithms%
  \thanks{HB is supported by a Vici grant from the Netherlands Organization for Scientific Research (NWO). RdW is supported by a Veni grant from NWO. HB, HR, and RdW are also supported in part by the EU fifth
    framework projects QAIP, IST-1999-11234, RESQ, IST-2001-37559, and the sizth framework project QAP.
    IN is partially supported by ISF grant 55/0.}} 
\author{Harry Buhrman\thanks{CWI, Amsterdam, the Netherlands and ILLC, University of Amsterdam, the Netherlands} 
  \and 
  Ilan Newman\thanks{Dept.~of Computer Science, Haifa University, Israel}
  \and  
  Hein R\"ohrig\thanks{Dept.~of Computer Science, University of Calgary, Canada}
  \and
  Ronald de Wolf\thanks{CWI, Amsterdam, the Netherlands}}
\date{}
\maketitle

\begin{abstract}
We define and study the complexity of \emph{robust} polynomials for Boolean
functions and the related fault-tolerant quantum decision trees, 
where input bits are perturbed by noise. We compare several different possible definitions.
Our main results are
\begin{itemize}
\item For every $n$-bit Boolean function $f$ there is an $n$-variate
polynomial $p$ of degree $\bigO(n)$ that \emph{robustly} approximates it,
in the sense that $p(x)$ remains close to $f(x)$ if we slightly vary each of 
the $n$ inputs of the polynomial.
\item There is an $\bigO(n)$-query quantum algorithm that \emph{robustly}
recovers $n$ noisy input bits.
Hence every $n$-bit function can be quantum computed with $\bigO(n)$
queries in the presence of noise.
This contrasts with the classical model of Feige~\etal, 
where functions such as parity need $\Theta(n\log n)$ queries.
\end{itemize}
We give several extensions and applications of these results.
\end{abstract}

\section{Introduction}\label{sec:robust-intro}

In the last two decades, polynomials of many varieties have been used 
quite successfully in complexity theory, both for upper and lower bounds. 
We study a variety here that is tailored to analyzing algorithms with \emph{noisy input}.

\paragraph{Robust Polynomials.}

A \emph{robust} polynomial for a Boolean function
$f:\01^n\rightarrow\01$ is a real multivariate polynomial
$p(z_1,\ldots,z_n)$ such that for every $x=(x_1,\ldots,x_n) \in
\{0,1\}^n$ and every $z=(z_1,\ldots,z_n) \in \R^n,$ if $\forall i :
|x_i - z_i| \leq 1/3$ then $|f(x) - p(z)| \leq 1/3$ (the $1/3$ in both
cases can be changed to any other positive constant less than 1/2). 
The \emph{robust degree}
of $f$ is the smallest degree of a robust polynomial for $f$; note
that we do not require robust polynomials to be multilinear.

The motivation behind the 
definition of robust polynomials is twofold. First, it can be
viewed as a strengthening (restriction) of the notion of approximating polynomials. 
An approximating polynomial for $f$ is a multivariate real polynomial
$q$ that approximates $f$ within an additive term of $1/3$ for each
Boolean input. Approximating polynomials for Boolean functions are
of interest in themselves and have been the object of study for quite a while. 
Their minimal degree is tightly related to the decision tree complexity of $f$
\cite{nisan&szegedy:degree,buhrman&wolf:dectreesurvey}. 
Indeed, this ``polynomial
method'' \cite{buhrman&wolf:dectreesurvey} is 
one of the main tools for obtaining lower bounds on the
number of queries in \emph{quantum} algorithms. One difficulty, however,
is that approximating polynomials do not directly compose: if
$f(x_1,\ldots,x_n)$ is a Boolean function with an approximating polynomial
$p_f$ and $g(y_1,\ldots,y_m)$ is a Boolean function with an approximating
polynomial $p_g$, then the polynomial on $n \cdot m$ variables
$p_f(p_g,\ldots,p_g)$ that is obtained by plugging in a copy of $p_g$ for
each of the $x_i$, is not necessarily an approximating
polynomial for the composed function $f(g,\ldots,g)$ on $n \cdot m$ variables.  
This difficulty is avoided with 
robust polynomials: if
$p_f, p_g$ are robust for $f,g$ respectively, then their composition is
a robust polynomial (and thus also approximating) for the composed
function.

A second motivation for robust polynomials is the study of 
quantum decision trees that can tolerate noise in their inputs. 
We show that a natural quantum analogue of classical fault-tolerant
decision trees can be defined. As a result, it will follow
that every such algorithm that uses $T$ queries to its input bits 
(and hence every classical noisy decision tree algorithm as well) 
implies the existence of a robust degree-$2T$ polynomial for the function. This
relates the robust degree to fault-tolerant quantum query algorithms
in exactly the same way that approximating polynomials are related to 
bounded-error quantum query algorithms. Surprisingly, our results 
imply robust quantum algorithms with a linear number of queries,
as well as robust polynomials of linear degree,  
for \emph{any} Boolean
function.
 This should be contrasted with the result of
 Feige~\etal~\cite{feige&al:computingWithNoisyInformation}. They proved
 that for most Boolean functions, an overhead factor of $\Omega(\log n)$
 on the number of queries is needed in the noisy case compared to the non-noisy
 case. In particular, consider the
 parity function on $n$ variables. This function can be decided
 trivially by an $n$-query decision tree, and hence can be represented
 exactly by a real multilinear polynomial of degree $n$ (which is just
 the single monomial containing all variables in the $\{-1,1\}$
 representation). Feige~\etal~\cite{feige&al:computingWithNoisyInformation}
prove that in the noisy decision tree model any algorithm for \prbl{Parity}
needs $\Theta (n \log n)$ queries. Using standard amplification techniques, 
this yields an $\bigO (n \log n)$-degree robust
polynomial for \prbl{Parity}. Can one do better?
 Our results imply that there is a robust polynomial for \prbl{Parity} of
 degree $\bigO(n)$. However, we only have an indirect description of
 this polynomial by means of a quantum algorithm, and do not know of
 an explicit 
 simple construction of such a polynomial.

\paragraph{Noisy Quantum Queries.}

We now discuss in more detail the model of noisy decision trees
in the quantum world.
The notion of a ``noisy query''  in
the quantum case is not as obvious and natural as in the classical case,
because one application of a quantum query 
can address many different $x_i$'s in superposition. 
A first proposal would be that for each quantum query, 
each of the bits is flipped independently with probability 
$\epsilon$. Each such quantum query
introduces a lot of randomness and the algorithm's state 
after the query is a mixed quantum state rather than a pure state.  
In fact, this model is a concrete (and very destructive) form of decoherence; the effects 
of various forms of decoherence on oracle algorithms like Grover's
have been studied before, see \eg \cite{long00:imperfectionGrover,shenvi03:noisyOracleGrover}.

A second model, which we will adopt here, is to assume
that we have $n$ quantum procedures, $A_1,\ldots,A_n$, such
that $A_i$ outputs $x_i$ with probability at least $1-\epsilon$.
Such a \emph{coherent-noise model}\/ is not unreasonable. 
For instance, it could be the case 
that the input bits are actually computed for us by 
subroutines. Such algorithms can always be made coherent by 
pushing measurements to the end, which means that we can apply 
and reverse them at will. To enable us to apply the $A_i$'s
in superposition, we assume we have a black box that maps
$$
  \mathcal A : \ket{i}\ket{0}\mapsto\ket{i}A_i\ket{0}. 
$$
One application of this will count as one query.

A third model, which we will call the \emph{multiple-noisy-copies model},
was studied by Szegedy and Chen~\cite{szegedy&chen:faulty}.
Here, instead of $x_i$, the algorithm can only query ``perturbed'' 
copies $y_{i,1}$, \ldots, $y_{i,m}$ of $x_i$. 
The $y_{i,j}$ are independent Boolean random variables with $\Pr [ x_i =
  y_{i,j} ] \ge 1 - \epsilon$ for each $i = 1, \ldots, n$ and $j =  1,\ldots,m$.
In contrast to the first proposal, this model leaves the 
queries perfectly reversible, since the perturbed copies 
are fixed at the start of the algorithm and the same 
$y_{i,j}$ can be queried more than once. 
The assumption of this model is also stronger than the second model,
since we can construct a 1-query $A_i$ that just outputs 
a superposition of all $y_{i,j}$. If $m$ is sufficiently large,
this $A_i$ will compute $x_i$ with high success probability,
satisfying the assumption of the second model
(see Section~\ref{subsec2models} for details).

\paragraph{Robust Quantum Algorithms.}

Assuming the second model of noisy queries and some fixed $\epsilon$, 
we call a quantum algorithm \emph{robust} if it computes $f$ on $n$ inputs 
with bounded error probability when the $n$ inputs are given by bounded-error 
algorithms $A_1,\ldots,A_n$, respectively. 

A first observation is that every 
$T$-query non-robust algorithm can be made robust at a multiplicative 
cost of $\bigO(\log T)$. With $\bigO ( \log T )$ queries,
a majority gate, and an uncomputation step, we can construct a unitary
$\tilde U_x$ that approximates an exact quantum query
$$
U_x : \ket i \ket b \mapsto \ket i \ket{ b \oplus x_i }
$$
very well in the standard operator norm: 
$\lb{ U_x - \tilde U_x } \le 1/(100T)$.  Since errors add
linearly in a quantum algorithm~\cite{bernstein&vazirani:qcomplexity}, 
replacing $U_x$ by $\tilde U_x$ in a
non-robust algorithm gives a robust algorithm with almost the same
final state.  In some cases better constructions are possible.  
For instance, a recent result by 
H\o{}yer~\etal~\cite{hmw:berrorsearch} implies a quantum
algorithm that robustly computes the $n$-bit \prbl{Or} function
with $\bigO(\sqrt{n})$ queries.
This is only a constant factor worse than the noiseless case,
which is Grover's algorithm~\cite{grover:search}.
In fact, we do not know of any function where the robust 
quantum query complexity is more than a constant factor larger 
than the non-robust complexity.

Our main result about robust quantum algorithms 
(made precise in Theorem~\ref{thm:recoverinput}) is the following:
\begin{quote}
There exists a quantum algorithm that outputs $x_1,\ldots,x_n$, with high probability,
using $\bigO(n)$ invocations of the $A_i$ algorithms (\ie queries).
\end{quote}
As already mentioned, this result 
implies that \emph{every} $n$-bit function $f$ can be
robustly quantum computed with $\bigO(n)$ queries. This contrasts 
with the classical $\Omega(n\log n)$ lower bound for \prbl{Parity}.
It is quite interesting to note that quantum computers,
which usually are more fragile than classical computers, are actually
more robust in the case of computing \prbl{Parity} in this model with noisy inputs.
The result for \prbl{Parity} can be extended to every symmetric 
function: for every such function, the optimal quantum
algorithm can be made robust with only a constant factor 
overhead (see Section~\ref{subsecconsequences}).

Our result has a direct bearing on the \emph{direct-sum problem},
which is the question how the complexity of computing $n$
independent instances of a function scales with the complexity 
of one instance. 
One would expect that computing $n$ instances with bounded-error
takes no more than $n$ times the complexity of one instance.
However, since we want all $n$ instances to be computed correctly 
\emph{simultaneously} with high probability, the only known general 
method in the classical world
is to compute each instance with error probability reduced 
to $\bigO(1/n)$. This costs another factor of $\bigO(\log n)$.
In fact, it follows from the $\Omega(n\log n)$ bound for \prbl{Parity} 
that this factor of $\log n$ is optimal if we can 
only run algorithms for individual instances in a black-box fashion.
In contrast, our result implies that in the quantum world,
the bounded-error complexity of $n$ instances is at most $\bigO(n)$
times the bounded-error complexity of one instance.  
This is a very general result. For example, it also applies
to communication complexity~\cite[Section 4.1.1]{kushilevitz&nisan:cc}.
If Alice and Bob have a bounded-error protocol for a distributed 
function $f$, using $c$ bits (or qubits) of communication, 
then there is a bounded-error quantum protocol for $n$ instances 
of $f$, using $\bigO(n(c+\log n))$ qubits of communication.  
The additive $\log n$ is because Alice and Bob need to 
communicate (possibly in superposition) the index 
of the instance that they are computing.
In contrast, the best known general classical solution 
uses $\Theta(cn\log n)$ bits of communication.

\paragraph{Note about Related Work.}

In their manuscript~\cite{ipy:biasedoracle}, 
Iwama~\etal study a similar but slightly weaker setting.
There, the error probability for each input variable is \emph{exactly} 
$\epsilon$.
If $\epsilon$ is known, then one can use a version of exact 
amplitude amplification to ``rotate off'' the error using $\bigO(1)$
queries and hence make the algorithm robust. 
If $\epsilon$ unknown, it can be estimated very well using quantum 
amplitude estimation, after which amplitude amplification 
can be used as if $\epsilon$ was known.
Iwama~\etal derive from this that any quantum algorithm 
can be made robust (in their model) with only a constant factor overhead.
Their model has the disadvantage that it does not cover the subroutine-scenario,
where each input bit $x_i$ is computed for us by an algorithm or subroutine
$A_i$ whose error we can only upper bound. Our model does not need the assumption
that the error is the same for all input bits, and hence does not have this
disadvantage.

\section{Robust Polynomials --- Preliminaries}
\label{sec:polynomials}

In this section we study robust polynomials of two different but
essentially equivalent types. The first type arises from the
multiple-noisy-copies model, the second type is what we discussed in
the introduction. 

\subsection{Two definitions}

\begin{definition}
Let $\epsilon\in[0,1/2)$.
  An \emph{$(\epsilon,m)$-perturbation} of $x \in
  \01^n$ is a matrix $y$ of $n \times m$ independent binary random
  variables $y_{i,j}$ such that $\Pr [ y_{i,j} = x_i ] \ge 1 - \epsilon$
  for each $1 \le j \le m$.
\end{definition}
\begin{definition}\label{def:typeone}
  A \emph{type-1 $(\epsilon,m)$-robust polynomial} for the Boolean
  function $f : \01^n \rightarrow \01$ is a real polynomial $p$ in $n m$
  variables $y_{i,j}$ (with $1 \le i \le n$ and $1 \le j \le m$) so
  that for every $x \in \01^n$ and $y$ an $(\epsilon, m)$-perturbation of
  $x$, we have 
$$
\Pr [ |p(y) - f(x)| > 1/3 ] <   1/3,
$$ 
where the probability is taken over the distribution on the $nm$ bits in $y$. 
Moreover, for every $v \in \01^{nm}$, we require $-1/3 \le p(v) \le 4/3$.
\end{definition}
Since $y_{i,j}^2=y_{i,j}$ for a bit $y_{i,j}$,
we can restrict attention to \emph{multilinear} polynomials here.

Notice that the error parameter $1/3$ in our definition of type-1 
polynomial is
consistent with having \emph{expected} error more than $1/2$ for some $x$:
it could be that $|p(y)-f(x)|=1/3$ with probability $2/3$, and 
$|p(y)-f(x)|=4/3$ with probability $1/3$, giving expected error $2/3$.
However, this is not a significant problem, as the next lemma shows that
the error parameter $1/3$ can be reduced to any small $\delta>0$ at only 
a small multiplicative cost in the degree and the number of perturbations.
It employs the following Chernoff bound from~\cite[Theorem~A.1.16]{alon&spencer:probmethod}.

\begin{theorem}[Chernoff]
Let $X_i$, $1\leq i\leq k$, be mutually independent random variables with all $\expect [X_i]=0$ and all $|X_i|\leq 1$. 
Set $S=\sum_{i=1}^k X_i$.
Then $\Pr[S>a]\leq e^{-a^2/2k}$.
\end{theorem}

\begin{lemma}\label{lem:typeoneboost}
Consider any $\delta>0$. 
If there is a \emph{type-1 $(\epsilon,m)$-robust polynomial} $p$ for $f$ 
of degree $d$, then there exists a type-1 $(\epsilon, m')$-robust polynomial $q$ 
for $f$ of degree $\bigO(d\log(1/\delta))$ and $m'=\bigO(m\log(1/\delta))$,
such that for $x \in \01^n$ and $y$ an $(\epsilon, m')$-perturbation of $x$,
we have 
$$
\Pr [ |q(y) - f(x)| > \delta] < \delta. 
$$
Moreover, for every $v \in\01^{nm'}$ we have $q(v)\in[0,1]$.
\end{lemma}

\begin{proof}
We first analyze the following single-variate ``amplification polynomial'' of degree $k$:
$$
h_k(x)=\sum_{i>k/2}{k\choose i}x^i(1-x)^{k-i}.
$$
Note that $h_k(x)$ is exactly the probability that among $k$ coin flips 
with bias $x$ towards 1, more than half come up 1.
Since it's a probability, we have $h_k(x)\in[0,1]$ for all $x\in[0,1]$. 
Moreover, applying the Chernoff bound with $X_i$ being the outcome of the $i$th coin flip minus $x$,
and $a=(1/2-x)k$, we have $h_k(x)\in[0,2^{-\Omega(k)}]$ for all
$x\in[0,1/3]$. Similarly $h_k(x)\in[1-2^{-\Omega(k)},1]$ for $x\in[2/3,1]$.
By ``stretching'' the domain a bit, we can turn this into a degree-$k$
polynomial $h_k$ such that $h_k(x)\in[0,2^{-\Omega(k)}]$ for $x\in[-2/5,2/5]$,
$h_k(x)\in[0,1]$ for $x\in[2/5,3/5]$, and
$h_k(x)\in[1-2^{-\Omega(k)},1]$ for $x\in[3/5,7/5]$.

We use $r$ independent $(\epsilon,m)$-perturbations of $x$, 
denoted $y=y_1,\ldots,y_r$, for some number $r$ to be determined later. 
For each perturbation $y_i$ it holds that $\Pr [ |p(y_i) - f(x)| > 1/3 ] < 1/3$. 
Using the amplification polynomial $h_k$ with $k=\bigO(1)$
we can get the value of $p$ closer to $f$:
$\Pr [ |h_k(p(y_i)) - f(x)| > 1/20 ] < 1/3$.
Note that the expected value of $|h_k(p(y_i)) - f(x)|$ is now at most
$(2/3)(1/20)+(1/3)1=11/30$. Now define an average polynomial
$\overline{p}(y)=\frac{1}{r}\sum_{i=1}^r h_k(p(y_i))$. 
Choosing $r=\bigO(\log(1/\delta))$, the Chernoff bound (with $k=r$, and $X_i$ being the indicator random variable for the event that 
$|h_k(p(y_i)) - f(x)|>23/60$ minus its expectation) we have 
$$
\Pr [ |\overline{p}(y) - f(x)| > 2/5 ] < \delta.
$$
Finally we apply $h_k$ again, this time with degree $k=\bigO(\log(1/\delta))$, 
in order to get the value of $\overline{p}$ $\delta$-close to the value $f(x)$:
if we define $q(y)=h_k(\overline{p}(y))$ then 
$$
\Pr [ |q(y) - f(x)| > \delta ] < \delta.
$$
The degree of $q$ is $\bigO(d\log(1/\delta))$, and $m'=mr=\bigO(m\log(1/\delta))$.
The last property of the lemma is also easily seen.
\qed
\end{proof}

The second kind of robust polynomial is the following:
\begin{definition}\label{def:typetwo}
  For a Boolean function $f : \01^n \rightarrow \01$, we call $q$ a
  \emph{type-2 $\epsilon$-robust polynomial for $f$} if $q$ is a real
  polynomial in $n$ variables such that for every $x \in \01^n$ and
  every $z \in [0,1]^n$ we have $| q(z) - f(x) | \le 1/3$ if $| z_i -
  x_i | \le \epsilon$ for all $i \in [n]$. If $\epsilon=0$, then $q$
  is called an \emph{approximating polynomial} for $f$.
\end{definition}
Note that we restrict the $z_i$'s to lie in the set $[0,\epsilon]\cup[1-\epsilon,1]$
rather than the less restrictive $[-\epsilon,\epsilon]\cup[1-\epsilon,1+\epsilon]$.
This facilitates later proofs, because it enables us to interpret
the $z_i$'s as probabilities.  However, with some extra work
we could also use the less restrictive definition here.
Also note that a minimal-degree type-2 robust polynomial for $f$
need not be multilinear, in contrast to the type-1 variety.

\begin{definition}
  For $f : \01^n \rightarrow \01$, let\/ $\robustdeg_1 ( f )$ denote the
  minimum degree of any type-1 $(1/3,\bigO(\log n)$-robust polynomial for $f$,
  let $\robustdeg_2 ( f )$ be the minimum degree of any type-2
  $1/3$-robust polynomial for $f$, and let $\approxdeg(f)$ 
  be the minimum degree among all approximating polynomials for $f$. 
\end{definition}

Strictly speaking, we should fix an explicit constant for the $\bigO(\log n)$
of the type-1 polynomial, but to simplify proofs we'll use the $\bigO(\cdot)$ instead.

\subsection{Relation between type-1 and type-2 robust polynomials}

We characterize the relation of type-1 and type-2
robust polynomials as follows:
\begin{theorem}\label{thm:typeonetwoequiv}
  For every type-2 $\epsilon$-robust polynomial of degree $d$ for $f$ there is a type-1
  $(\epsilon/2,\bigO(\log(n)/\linebreak[2](1/2-\epsilon)^2))$-robust polynomial of
  degree $d$ for $f$.  

Conversely, for every type-1 ($\epsilon,m)$-robust polynomial of degree $d$ for $f$ there is a
  type-2 $\epsilon$-robust polynomial of degree $\bigO(d)$ for $f$.
\end{theorem}

\begin{proof}
Let $p$ be a type-2 $\epsilon$-robust polynomial of degree $d$ for $f$.
We choose $m=\bigO(\log(n)/(1/2-\epsilon)^2)$.
If each $y_{i,j}$ is wrong with probability $\leq \epsilon/2$, 
then the Chernoff bound implies that the probability that the average 
$\overline{y}_i=\sum_{j=1}^m y_{i,j}/m$ is more than $\epsilon$ away from $x_i$, is at most $1/(3n)$.
Then by the union bound, with probability at least $2/3$ we have $|\overline{y}_i-x_i|\leq \epsilon$ 
for all $i\in[n]$ simultaneously.  Hence the polynomial $p(\overline{y}_1,\ldots,\overline{y}_n)$
will be a type-1 $(\epsilon/2,\bigO(\log(n)/(1/2-\epsilon)^2))$-robust polynomial of
degree $d$ for $f$.
  
  For the other direction, consider a type-1 $(\epsilon,  m)$-robust 
  polynomial of degree $d$ for $f$. Using Lemma~\ref{lem:typeoneboost},
  we boost the approximation parameters to obtain a type-1 $(\epsilon,
  m')$-robust polynomial $p$ of degree $\bigO(d)$, with $m'=\bigO(m)$, 
  such that for any $x \in \01^n$ and 
  $(\epsilon, m')$-perturbation $y$ of $x$, $\Pr [ |p(y) - f(x)| > 1/9 ]
  < 1/9$. For $z=(z_1, \ldots , z_n)$ define the formal polynomial
  $q(z)$ (over the reals) by replacing each appearance of $y_{i,j}$
  in $p(y)$ with $z_i$. For $z \in \R^n$ with $0 \le z_i \le
  1$ for all $i$, let $y_{i,j}$ ($i \in [n]$, $j \in [m']$) be
  independent $0/1$ random variables, where $\expect [ y_{i,j} ] =
  z_i$.
 Then the polynomial $q(z)$ that is defined above can be
  viewed as $q(z) = \expect [ p(y) ]$
  because $\expect [ p(y) ] = p ( \expect [ y ] )$ and 
  $\expect [ y_{i,j} ] = z_i$. In particular, if for $z$ there exists $x \in \01^n$
  with $|z_i - x_i| \le \epsilon$ for all $i$, then for any
   $y\in \{0,1\}^{nm}$ that is an $(\epsilon,m)$-perturbation of $x$, we have 
 $q(z) = \expect [ p(y) ]$ (here expectation is according to the
 distribution induced by $y$). 
  Therefore $V := \{ v\in \{0,1\}^{nm} : |p(v) -
  f(x) | < 1/9 \}$ has probability $\Pr[ y \in V ] > 8/9$ and
  \begin{equation*}
    \left| f(x) - q(z) \right| \le \left| \sum_{v \in V} \Pr[y = v] \left( f(x) - p(v) \right)
    \right| + \left| \sum_{v \notin V} \Pr[y = v] \left( 1+\frac19
      \right) \right| < \frac13 \enspace .
  \end{equation*}
  This means that $q(z)$ is a type-2 $\epsilon$-robust polynomial for
  $f$ of degree $\bigO( d )$.
\qed
\end{proof}

Note, in all the above we have discussed total Boolean functions. The
definitions above make sense also for partial Boolean functions (or
promise problems). The theorem as well as the next corollary are true
also for such cases.
\ignore{
Note that in Definition~\ref{def:typeone} we require for type-1
polynomials $p$ that for any Boolean assignment $v \in \01^{nm}$ to
the (possibly real) variables, the polynomial value $p(v)$ lies between
$-1/3$ and $4/3$. Because of this totality requirement, the following
corollary is given for total Boolean $f$ only.
}
\begin{corollary}\label{cor:deg1deg2Equiv}
  $\robustdeg_1 ( f ) = \Theta(\robustdeg_2 ( f ) )$ for every 
  Boolean function $f : \01^n \rightarrow \01$.
\end{corollary}

\subsection{Polynomials induced by quantum algorithms}

The well known ``polynomial method''~\cite{bbcmw:polynomials}
allows us to make a connection between ``robust'' quantum algorithms and 
robust type-1 polynomials:

\begin{lemma}\label{lem:functionToType1Poly}
  Let $f : \01^n \rightarrow \01$ be a Boolean function.  Let $Q$ be a
  quantum algorithm that makes at most $T$ queries on inputs $y$ from
  $\01^{n \times m}$, and let $Q(y)$ denote the binary random variable that is its output. 
   If for every $x \in \01^n$ and $y$ an
  $(\epsilon, m)$-perturbation of $x$, we have that $\Pr_y[ Q(y) = f(x) ] \ge 8/9$
  (probability taken over the distribution on the $nm$ bits in $y$ as well as over the algorithm),
  then there exists a degree-$2T$ type-1 $(\epsilon, m)$-robust polynomial for $f$.
\end{lemma}

\begin{proof}
  By \cite[Lemma~4.2]{bbcmw:polynomials}, $Q$ induces a degree-$2T$
  multilinear polynomial $p$ on $n m$ variables that gives the
  acceptance probability of $Q$ on fixed input $y \in \01^{nm}$, \ie
  $p(y) = \Pr [ Q(y) = 1]$ (probability taken only over the algorithm).  
Fix $x \in \01^n$. Suppose $f(x)=0$, then we want to show that $\Pr_y[p(y)>1/3]<1/3$.
Since $\Pr_y[ Q(y) = f(x) = 0 ] \ge 8/9$, we have $\expect_y[p(y)]=\Pr_y[Q(y)=1]\leq 1/9$. 
Hence Markov's inequality implies $\Pr_y[p(y)>1/3]<1/3$ and we are done.
The case $f(x)=1$ is similar.
\qed
\end{proof}

\section{Quantum Robust Input Recovery}\label{sec:recoverinput}

In this section we prove our main result, that we can recover an
$n$-bit string $x$ using $\bigO(n)$ invocations of algorithms
$A_1,\ldots,A_n$ where $A_i$ computes $x_i$ with bounded error.
Let $|x|$ denote the Hamming weight of a bit string $x$.
Our main theorem says that with high probability we can find 
$t$ 1-bits in the input $x$ (if they are present) using $\bigO(\sqrt{n t})$ noisy queries.

\begin{theorem}\label{thm:recoverinput}
  Let $\epsilon\in[0,1/2)$.
  Consider $\epsilon$-error algorithms $A_1$, \ldots, $A_n$ that
  compute the bits $x = x_1,\ldots,x_n$.  
For every $t$, $1 \le t
  \le n$, there is a quantum algorithm that makes $\bigO(\sqrt{n t}
)$ queries (invocations of the $A_i$) and that
  outputs $\tilde x = \tilde x_1,\ldots,\tilde x_n$ such that with
  probability at least 2/3
  \begin{enumerate}
  \item for all $i$: $\tilde x_i = 1 \Rightarrow x_i = 1$
  \item $|\tilde x| \ge \min \{ t, |x| \}$.
  \end{enumerate}
\end{theorem}
In particular, with $t = n$ we obtain $\tilde x = x$ using
$\bigO(n)$
queries.

\subsection{Some more preliminaries}

For simplicity we assume that $ 0 <\epsilon < 1/100$ is fixed and that $A_i$ 
is a unitary transformation
\[
  A_i : \ket{0^t} \mapsto \alpha_i \ket 0 \ket{\psi^0_i} + 
  \sqrt{1-\alpha_i^2} \ket1 \ket{\psi^1_i}
\]
for some $\alpha_i\geq 0$ such that $|\alpha_i|^2 \le
\epsilon$ if $x_i=1$ and $|\alpha_i|^2 \ge 1-\epsilon$ if $x_i=0$;
$\ket{\psi^0_i}$ and $\ket{\psi^1_i}$ are arbitrary norm-1 quantum states. 
The \emph{output} is the random variable obtained from measuring the first qubit.
It equals $x_i$ with probability at least $1-\epsilon$.
It is standard that any quantum algorithm can be expressed in this form by postponing
measurements (\ie unitarily write the measurement in an auxiliary 
register without collapsing the state); any classical randomized algorithm can 
be converted into this form by making it reversible and replacing random bits by states
$(\ket0+\ket1)/\sqrt2$. 

We define the following notion of closeness:

\begin{definition}
  For $\epsilon\in[0,1/2)$ and algorithms $\mathcal A = ( A_1,
  \ldots, A_n ) $, we say $\mathcal A$ is \emph{$\epsilon$-close} to $x
  \in \01^n$ if $\Pr[ A_i \mbox{ outputs } x_i ] \ge 1-\epsilon$ for all $i \in [n]$.
\end{definition}
%
We sometimes modify our sequence of algorithms $\mathcal A = ( A_1, \ldots, A_n ) $ as follows.
For an $n$-bit string $\widetilde{x}$, we negate the answer of $A_i$ if $\widetilde{x}_i=1$,
and denote the resulting sequence of $n$ algorithms by $\mathcal A(\widetilde{x})$.
Note that $\mathcal A(\widetilde{x})$ is close to $0^n$ if and only if $\widetilde{x}=x$.
In other words, by finding ones in $\mathcal A(\widetilde{x})$, 
we find positions where $\widetilde{x}$ differs from $x$.
In addition, for a set $S\subseteq [n]$ we use ${\mathcal A}^S(\widetilde{x})$ 
to denote the vector of algorithms ${\mathcal A}(\widetilde{x})$,
except that for all $i\not\in S$ the $i$th algorithm always outputs 0 instead of running $A_i$.
Also, for $S$ as above and $x \in \{0,1\}^n$ we denote by $x^S \in \{0,1\}^n$ the string that is
identical to $x$ on indices in $S$ and is $0$ on indices in $\bar{S}$.

Our algorithm builds on a robust quantum search algorithm by H{\o}yer,
Mosca, and de~Wolf~\cite{hmw:berrorsearch}, which we call \procref{RobustFind}.
This subroutine takes a vector $\mathcal
A$ of $n$ quantum algorithms and in the good case returns an index $i$
such that the ``high probability'' output of $A_i$ is $1$. Formally, the
input/output relation of \procref{RobustFind} is stated in Theorem
\ref{thm:robustfind}. 
\begin{theorem}[H\o yer, Mosca, de Wolf~\cite{hmw:berrorsearch}]\label{thm:robustfind}
There is a procedure {\bf RobustFind($n$, $\mathcal A$, $\epsilon$, 
    $\beta$, $\gamma$, $\delta$)} where 
  $n \in \N$, $\mathcal A$ : $n$ quantum algorithms,
  $\epsilon,\beta,\gamma,\delta>0$\\
    {\bf Output: }
    $i \in  [n] \cup \{ \perp \} $ and with the following properties:
    \begin{enumerate}
    \item if $\mathcal A$ is $\epsilon$-close to $x \in \01^n$ and $x$ has Hamming weight $|x|
      \ge \beta n$, then $i \ne \perp$ with probability 
      $ \ge 1-\delta$
    \item if $\mathcal A$ is $\epsilon$-close to $x \in \01^n$ and if $i
      \ne \perp$, then $x_i = 1$ with probability $\ge 1-\gamma$
    \end{enumerate}
    {\bf Complexity: }
    $ \bigO \left( 
      \frac1{\left( \frac12 - \epsilon \right)^2 } \cdot  \sqrt{ \frac1\beta} \cdot \log\frac1{ \gamma \delta} \right)
    \text{ invocations of the } A_i $
\end{theorem}

\subsection{The algorithm and its intuition}

Before we formally prove Theorem \ref{thm:recoverinput} we explain the
intuition and high level of our algorithm (as defined by the 
\procref{AllInputs} pseudo code
on page~\pageref{proc:AllInputs}) and of the proof.
Clearly, for  $t=\bigO(1)$  Theorem \ref{thm:recoverinput} is obvious as 
we can run \procref{RobustFind} $t$ times to
recover $t$ indices $i$ such that $x_i = 1$ with $\bigO (
\sqrt{n} )$ queries. Therefore all
considerations below will be for $t > t_0$ for some $t_0$ that is
independent of $n$ and will be specified later.

An important feature of the robust quantum search is that it can be
used to verify a purported solution $\tilde x \in \01^n$ by running
\procref{RobustFind} on $\mathcal A (\tilde x)$ to find differences with the
real input $x$.

\begin{algorithm}[htb]
  \procdef{AllInputs}
  \caption{AllInputs($n$, $t$, $\mathcal A$, $\epsilon$)}
  $n, t \in \N$, $\mathcal A$ : $n$ algorithms, $\epsilon > 0$
  \begin{algorithmic}[1]
    \STATE $\tilde{x}\leftarrow 0^n$ 

\vspace{0.7cm}
{\bf Part 1, Aim: } {\it to find a set of indices } $S \subseteq [n]$ {\it that contains
at least} $\min(|x|,t)$ {\it and at most} $3t/2$ $1$'s {\it of the
input}.
\vspace{0.1cm}
    \FOR{$3t/2$ times}
    \label{alg:Begin:InitialGuess}
    \STATE $i \leftarrow \procref{RobustFind} (n, \mathcal
    A(\tilde{x}), \epsilon, \frac {t}{100n}, \frac{1}{100}, \frac{1}{100} )$
    \IF{ $i \ne \perp$}
    \STATE $\tilde{x}_i \leftarrow 1 - \tilde{x}_i$
    \label{alg:End:InitialGuess}
    \ENDIF
    \ENDFOR

   \STATE $S\leftarrow\{i \mid \tilde{x}_i =1 \}$
  \IF{ $|S|< 5t/4$} 
  \STATE  $S \leftarrow [n]$
  \ENDIF
\vspace{0.7cm}

{\bf Part 2, Aim: } {\it correctly find all but $t/\log^2 t$ $1$'s}.

 \STATE $\beta \leftarrow \frac{t}{100n}$
 \STATE $\tilde{x} \leftarrow 0^n$
    \FOR{$k \leftarrow 1$ to $\log ((\log t)^2 )$ }
    \label{alg:Begin:SampleBad}
    \STATE $\beta_k \leftarrow \beta /2^{k}$
    \STATE $t_k \leftarrow 3t/ 2^{k}$
    \FOR{$\ell \leftarrow 1$ to $ t_k$}
    \STATE \label{alg:bkl}$i \leftarrow \procref{RobustFind} 
    (n, \mathcal A^S(\tilde{x}), \epsilon, \beta_k n, \frac{1}{100}, \frac{1}{100} )$
    \IF{ $i \ne \perp$}
    \STATE $\tilde{x}_i \leftarrow 1 - \tilde{x}_i$
    \label{alg:End:SampleBad}
    \ENDIF
    \ENDFOR
    \ENDFOR 

\vspace{0.7cm}
{\bf Part 3, Aim: } {\it correctly find all other $1$'s and get rid of
  remaining errors.}

    \FOR{$m \leftarrow t/(\log t)^2$ down to $1$}
    \label{alg:Begin:FindAllBad}
    \STATE $i \leftarrow \procref{RobustFind} (n, \mathcal A^S(\tilde{x}), 
    \epsilon, \frac{m}{n}, \frac1{20t}, \frac1{20t})$
    \IF{ $i \ne \perp$}
    \STATE $\tilde{x}_i \leftarrow 1 - \tilde{x}_i$
    \label{alg:End:FindAllBad}
    \ENDIF
    \ENDFOR    
    \STATE {\bf
      return} $\tilde{x}$
  \end{algorithmic}
\end{algorithm}

Let $x$ be the unique assignment such that $\mathcal{A}$ is
$\epsilon$-close to $x$. Assume first that the Hamming weight is $|x| <
3t/2$. 
 Our idea is to apply \procref{RobustFind} repeatedly  for about
$3t/2$ times (with threshold, say, $\beta = t/(100 n)$) and error
probability $1/100$. We expect that for at least a $98/100$-fraction of
the calls, \procref{RobustFind} will return an index $i$ such that
$x_i=1$, and expect at most a $2/100$-fraction of wrong indices. The
first problem to note is that \procref{RobustFind} might return the same
(correct) index
over and over again. This is easily resolved as follows: We set $\tilde{x} \in
\{0,1\}^n$ to be $\tilde{x}_i = 1$ for every index $i$ that we
obtained from \procref{RobustFind} and $0$ everywhere else, and call
\procref{RobustFind} with $\mathcal{A}(\tilde{x})$ rather than with
$\mathcal{A}$. This means that the $1$'s that are to be reported by
\procref{RobustFind} are in $x \xor \tilde{x}$ which is supported on the
erroneous indices of $\tilde{x}$, namely, on those indices that are
either $1$ in $\tilde{x}$ but are $0$ in $x$ (false positive) and
those indices that are $0$ on $\tilde{x}$ while they are $1$ on $x$
(false negative).

Done this, we expect about  $3t/200$
errors of both kinds (false positive and false negative) in the
$3t/2$ calls to \procref{RobustFind}, which should result in 
$\tilde{x}$ being quite close to $x$. We then 
call \procref{RobustFind} $3t/4$ times hoping to correct some of the
errors while not introducing too many new errors. This would be
reasonable as we call \procref{RobustFind} in this second phase half the
times we call it in the first phase. Thus we expect to have half the 
number of new errors, while good chance of correcting many old errors
(as they are $1$ in $x \xor \tilde{x}$ and hence \procref{RobustFind} is
expected to report a $98/100$-fraction of them). We keep doing this
until the number of expected errors is smaller than $t/(\log^2 t)$. At
this point we can afford to run \procref{RobustFind} for $t/(\log^2 t)$ times, 
with error probability as
low as $1/(20t)$. This finds all remaining errors with high
probability. Indeed this is the structure of part 2 and part 3 of our
algorithm. 

However, the idea above fails to work when $|x| \gg t$. To see the problem
assume that $t=\sqrt{n}$ while $|x|=n/2$. Then, after the first round
 above, $\tilde{x}$ will be supported on about
$\sqrt{n}$ indices, out of which about $\sqrt{n}/100$ might be false
positives. However, in every next call to \procref{RobustFind}, the procedure
has about $n/2 - \sqrt{n}$ false negative indices to report back  - those that
are $1$'s in $x$ but still $0$ in $\tilde{x}$. Thus, even if all the
next $\bigO(t)$ calls return a correct such index, we still might
be left with the same $\sqrt{n}/100$ false positive errors that are introduced in
the first round. Note that if $t= n$, which is the case when the
algorithm is applied to find all inputs, this last discussion is of no
concern. However, for relatively small $t$ (which will be needed for
some applications, e.g., Theorem ~\ref{thm:symmetric}) we need to introduce a first
part to the algorithm. This part is only meant to find a subset $S
\subseteq [n]$ such that $ |x^S| < 3t/2$. Once this is done, we can
use $x^S$ instead of $x$ in the description above, which will now work
for every input. 

\subsection{Detailed proof}

We now prove that the success probability of the algorithm is
at least $2/3$.
\paragraph{Success probability.}

The algorithm is composed of three parts. We first prove that after
Part~1, that is, prior to line~9,  we have
$\min(t, |x|) \leq |x^S| \leq 3t/2$ with probability $1-o(1)$. 

Indeed, assume first that just prior to the execution of line~7 we have $|S| \geq 5t/4$. Then the
upper bound on $|x^S|$ is trivial. For the lower bound assume (by way of
contradiction) that
$|x^S| < t$. Then we can have $|S| \geq 5t/4$ only if at least $t/4$
wrong indices have been reported by \procref{RobustFind}. However, as we call
\procref{RobustFind} with $\gamma = 1/100$ we expect at most $3t/200$
errors. Thus by the Chernoff bound we have $|x^S| \geq t$ with probability $1-o(1)$.

If, on the other hand, we reach line~7 with $|S| < 5t/4$ then 
 $S$ is set to be $[n]$, for which the lower bound on $|x^S|$ certainly
holds. For the upper bound assume that $|x| \geq 3t/2$. Then to have
$|S| < 5t/4$ at line~7 means that at least $t/4- t/100$ errors occurred
in the $3t/2$ calls for \procref{RobustFind} (an error here is whenever 
\procref{RobustFind} returns either  $i = \perp$ or a false negative index;
the $t/100$ term
comes from the threshold $\beta  = t/(100n)$). However, the error
probability in this case is at most $2/100$ (as we call \procref{RobustFind}
with $\delta= \gamma = 1/100$). Thus we expect at most $3t/100$
errors. Again by Chernoff we are done.

Accordingly, we may assume that with probability $1-o(1),$ the $S$ we have at
line~9 is such that $\min(t, |x|) \leq |x^S| \leq 3t/2$. 
In Part~2 of the algorithm we  want to find \emph{correctly} most of
the $1$'s in $x^S$. We maintain $\tilde{x}$ as our current estimate 
 of $x^S$. Initially $\tilde{x}=0^n$. Denote by $G_k, k=1, \ldots , \log((\log
t)^2)$ the event that $|\tilde{x} \xor x^S| <30t_k/100$ at the end of
the $k$th run of the loop in line~10;  $\bar{G}_k$ denotes the 
complementary event (the negation of $G_k$).
We prove inductively that
$\Pr[\bar{G}_k|G_{k-1}]  = e^{-\Omega(t_k)}$. This together with an
assertion  that $\Pr[G_1]  =  e^{-\Omega(t)}$ 
  will imply that at the end of Part~2, $|x^S \xor \tilde{x}| \leq t/\log^2
  t $ with probability at least $9/10$, assuming that $t$ is large enough
  (such that $ e^{-\Omega(t_k)} =  e^{-\Omega(t/\log^2 t)} < 1/(10\log(
  \log^2 t))$). 

Indeed, let us examine the situation during the first round, namely for $k=1$.
 We
 call \procref{RobustFind} in the first round for $t_1 = 3t/2$ times with threshold  $\beta_1 n =
 t/200$. Thus, as long as
$|x^S \xor \tilde{x}| > t/200$ happens, each call to \procref{RobustFind} gives 
an $i \in [n]$ with probability at least $99/100$. Moreover, we expect at most
a $1/100$ fraction of errors in the reported indices. Assume first that at the
beginning of the first round  $|x^S \xor \tilde{x}| > 20t/100$ 
and let $h=|x^S \xor \tilde{x}|- t/200$. Then after the first $h$
calls to \procref{RobustFind} we expect at least $98/100$ fraction of correct
indices.  Thus with probability $e^{-\Omega (t)}$ we will get less
than $90/100$ correct indices. However, if we do get at least
$\frac{90}{100} \cdot |x^S \xor \tilde{x}|$ correct indices after
  those $h$ calls we get an $\tilde{x}$ for which  $|x^S \xor \tilde{x}| \leq
  \frac{20}{100} h \leq  6t/100$. Now, assuming this happens, then
  $\bar{G}_1$ can happen at the end of the first round only if during the
  rest of the $3t/2 - h \leq 129t/100$ remaining calls  at least
  $39t/100$ incorrect indices have been made. As the probability for
  an incorrect index is bounded by $1/100$ we expect only at most
  $1.3t/100$ errors. Thus, by Chernoff $39t/100$ errors will occur
  with probability $e^{-\Omega (t)}$. If, however, at the beginning of
  the first round $|x^S \xor \tilde{x}|\leq 20t/100$ then by a similar
  argument 
$\bar{G}_1$ can happen at the end of the first round only if during 
  the $3t/2$  calls to \procref{RobustFind}  at least
  $25t/100$ incorrect indices have been made. Again by Chernoff this
  will happen with probability $ e^{-\Omega (t)}$. 
This concludes the proof that
  $\Pr[\bar{G}_1] = e^{-\Omega (t)}$.

We now inductively prove that $\Pr[\bar{G}_k | G_{k-1}] \leq e^{-\Omega(t_k)}$.

Indeed, assume that $G_{k-1}$ happens, namely that just prior to the beginning of
the $k$th round we have $|\tilde{x} \xor x^S| < 30t_{k-1}/100 =
60t_k/100$. In round $k$ we call
\procref{RobustFind} with threshold $\beta_k n = t_k/200$; 
hence, as long as  $|\tilde{x} \xor x^S| >t_k/200$, 
we expect \procref{RobustFind} to return an index $i \in [n]\cap S$
with probability at least $99/100$. Moreover, every time
it returns a correct index (which occurs with probability at least $99/100$)
it is a $1$ in
$(\tilde{x} \xor x^S)$ hence reduces the weight of symmetric difference
(the total number of errors) by $1$. 

Suppose first that prior to
round $k$, $|\tilde{x} \xor x^S| < 30t_k/100$. Then, for $\bar{G}_k$ to
happen at the end of round $k$,  \procref{RobustFind} would need to return at
least $31t_k/100$ wrong indices, namely $i \in [n] \cap S$ such that
$\tilde{x}_i = x_i$ (returning a $\perp$ here does not count as a
false index). However, as the probability of a wrong index is at most
$1/100$ and \procref{RobustFind} is called $t_k$ times, then, by Chernoff, the
probability of $\bar{G}_k$ is $e^{-\Omega(t_k)}$. 

Assume  now that  $|\tilde{x} \xor x^S| \geq
30t_k/100$ at the beginning of round $k$. 
Recall also that by the assumption that $G_{k-1}$
 occurs, we have $|\tilde{x} \xor x^S| <
60t_k/100$ at the beginning of the $k$th round. 
Consider the first $h=|\tilde{x} \xor x^S| - t_k/200$ calls
for \procref{RobustFind}. In each of those calls $|\tilde{x} \xor x^S| >
t_k/200= \beta_k n $, hence with probability $99/100$ every such call
returns an index $i \in [n] \cap S$ which is then a correct index
with probability $99/100$. Thus we expect that at least
$\frac{98}{100} \cdot h$ correct indices will be returned in the first
$h$ calls. By Chernoff, the
probability that the number of correctly returned indices in those
$h$ calls is less than $90h/100$ is  $e^{-\Omega(t_k)}$ (as $h \geq
15t_k/100$). But if the number of correctly returned indices is at
least $90h/100$, then after the first $h$ calls of
\procref{RobustFind}, $|\tilde{x} \xor x^S| < 0.2h \leq 0.2 \cdot 59t_k/100 <
12t_k/100$. Thus, at this point we are still left with $3t_k/2 -h$
calls to \procref{RobustFind} which will result in $\bar{G}_k$ only if at
least $48t_k/100$ wrong indices will be returned. This again will
happen with probability $e^{-\Omega(t_k)}$. We conclude that in all
cases $\Pr[\bar{G}_k|
G_{k-1}] = e^{-\Omega(t_k)}$.

Note that $t_k > t/(\log^2 t)$. Thus if we choose $t> t_0$ such that
for every $k$ the probability  $\Pr[\bar{G}_k|
G_{k-1}] = e^{-\Omega(t_k)} < 1/(10t)$ we get that $\Pr[\bar{G}_k] <
1/10$ for $k=\log^2 t$ after the end of Part~2. 
Hence,  with probability at least $0.8$, 
we have $|\tilde{x} \xor x| < t/(\log t)^2$ bad indices at the end of the 
\texttt{for} loop in lines~\ref{alg:Begin:SampleBad}--\ref{alg:End:SampleBad}. 

Finally, in Part~3 we  find (with probability close to 1) all remaining wrong
 indices by making the individual error probability
in \procref{RobustFind} so small that we can use the union bound: we
determine each of the remaining bad indices with error probability
$1/(10t)$.  This implies an overall success probability of at least 
$0.8\cdot 0.9 > 2/3$.

\paragraph{Complexity.}

Clearly the complexity is determined by Parts~2 and~3 of the algorithm.
We bound the number of queries to $f$ in lines~11--17 as follows:
\begin{equation}
\label{eq:compl}
\bigO\left(\sum_{k = 1}^{\log (\log^2 t)} t_k \sqrt{1/\beta_k}\right) = 
\bigO\left(\sum_{k = 1}^{\log (\log^2 t)}
  \frac{t}{2^k}\sqrt{ \frac {n 2^k} {t}}\right) = 
\bigO \left(
      \sqrt{n t} \right)
\end{equation}
The number of queries in lines~18--21 is bounded by
\begin{equation*}
  \bigO \left(
    \sum_{m = 1}^{t/(\log t)^2}
    \sqrt{ \frac n m }
    \log t
    \right)
    = \bigO \left(
\sqrt{n t}
\right).
\end{equation*}
Therefore, the total query complexity of
\procref{AllInputs} is $\bigO(\sqrt{n t}
)$. 

\section{Making Quantum Algorithms Robust}

\subsection{Inputs computed by quantum algorithms}\label{subsecconsequences}

Here we state a few corollaries of Theorem~\ref{thm:recoverinput}.
First, once we have recovered the input $x$ we can compute
any function of $x$ without further queries, hence
\begin{corollary}\label{cor:quantumRobustAtMostNQueries}
For every $f:\01^n\rightarrow\01$, there is a robust
quantum algorithm that computes $f$ using $\bigO(n)$ queries.
\end{corollary}
In particular, \prbl{Parity} can be robustly quantum computed with $\bigO(n)$ 
queries while it takes $\Omega(n\log n)$ queries 
classically~\cite{feige&al:computingWithNoisyInformation}.

Second, in the context of the direct-sum problem,
the complexity of quantum computing a vector of instances of
a function scales linearly with the complexity of one instance.
\begin{corollary}[Direct Sum]
If there exists a $T$-query bounded-error quantum algorithm for $f$,
then there is an $\bigO(Tn)$-query bounded-error quantum algorithm 
for $n$ independent instances of $f$.
\end{corollary}
As mentioned, the best classical upper bound has an additional factor 
of $\log n$, and this is optimal in a classical black-box setting.

Thirdly, all \emph{symmetric} functions can be computed robustly on a quantum
computer with the same asymptotic complexity as non-robustly. 
A function is symmetric if its value only depends on the Hamming 
weight of the input.
Let $\Gamma(f) := \min \{ |2k - n + 1| : f $ changes value if the Hamming
weight of the input changes from $k$ to $k+1 \}$. 
Beals~\etal~\cite[Theorem~4.10]{bbcmw:polynomials} exhibited a bounded-error 
quantum algorithm for $f$ using $ \bigO(\sqrt{n (n-\Gamma(f)+1)}) $ 
quantum queries, which is optimal.
We show that this upper bound remains valid also for \emph{robust} algorithms.

\begin{theorem}\label{thm:symmetric}
For every symmetric function $f$, there is a robust quantum algorithm 
that computes $f$ using $ \bigO(\sqrt{n (n-\Gamma(f)+1)}) $ quantum queries.
\end{theorem}

\begin{proof}
  Note that $f$ is constant when the Hamming weight of its input lies in 
  the middle interval $[(n-\Gamma(f))/2,(n+\Gamma(f)-2)/2]$.
  Using two applications of Theorem~\ref{thm:recoverinput} with sufficiently
  small error probability, we robustly search for $\lceil (n-\Gamma(f))/2\rceil$ ones 
  and $n - \lceil ( n + \Gamma(f) -2 ) /2 \rceil$ zeros in the input.
  If both of these searches succeeded (\ie found the required zeros and ones),
  then we know that our input lies in the middle interval. 
  If the search for zeros failed (\ie ended with fewer zeros) then we know 
  \emph{all}\/ zeros and hence the whole input $x$. Similarly, if the search 
  for ones failed then we know $x$.  Either way, we can output $f(x)$. \qed
\qed
\end{proof}

\subsection{Multiple noisy copies}\label{subsec2models}

As mentioned in the introduction, the assumption that we have
a bounded-error algorithm $A_i$ for each of the input bits
$x_i$ also covers the model of~\cite{szegedy&chen:faulty}
where we have a sequence $y_{i,1},\ldots,y_{i,m}$ 
of noisy copies of $x_i$.
These we can query by means of a mapping 
$$
\ket{i}\ket{j}\ket{0}\mapsto\ket{i}\ket{j}\ket{y_{i,j}}.
$$
Here we spell out this connection in some more detail.
First, by a Chernoff bound, choosing $m:=\bigO(\log(n)/\epsilon^2)$
implies that the average $\overline{y}_i:=\sum_{j=1}^m y_{i,j}/m$ 
is close to $x_i$ with very high probability:
$$
\Pr[|\overline{y}_i-x_i|\geq 2\epsilon]\leq \frac{1}{100n}.
$$
By the union bound, with probability $99/100$ this
closeness will hold for all $i\in[n]$ simultaneously.
Assuming this is the case, we implement the following unitary mapping
using one query:
$$
A_i:\ket{0^{\log(m)+1}}\mapsto\frac{1}{\sqrt{m}}\sum_{j=1}^m\ket{j}\ket{y_{i,j}}.
$$
Measuring the last qubit of the resulting state gives $x_i$ 
with probability at least $1-2\epsilon$. Hence, we can run our 
algorithm from Section~\ref{sec:recoverinput} and recover $x$ 
using $\bigO(n)$ queries to the $y_{i,j}$.
Similarly, all consequences mentioned in Section~\ref{subsecconsequences}
hold for this multiple-noisy-copies model as well.

\section{Making Approximating Polynomials Robust}

The next theorem follows immediately from earlier results.

\begin{theorem}
$\robustdeg_{1,2}(f)=\bigO(n)$ for every $f:\01^n\rightarrow\01$. 
\end{theorem}
\begin{proof}
  By Corollary~\ref{cor:quantumRobustAtMostNQueries} and the
  discussion in Section~\ref{subsec2models}, $f$ has an
  $\bigO(n)$-query robust quantum algorithm in the
  multiple-noisy-copies model that operates on $\bigO(\log n)$
  copies. By Lemma~\ref{lem:functionToType1Poly} this induces a type-1
  robust polynomial for $f$ of degree $\bigO(n)$. And finally, by
  Corollary~\ref{cor:deg1deg2Equiv} there also exists a
  degree-$\bigO(n)$ type-2 robust polynomial for $f$.
\qed
\end{proof}

In particular, this shows that for functions with approximate degree $\Theta(n)$
we can make the approximating polynomial robust at only constant factor overhead
in the degree. This case includes explicit functions like \prbl{Parity} 
and \prbl{Majority}, but also random (hence almost all) functions.
It is open whether approximating polynomials can \emph{always} be made
robust at only a constant overhead in the degree.
The best we can do is show that a non-robust degree-$d$ approximating polynomial
can be made robust at a cost of a factor $\bigO(\log d)$.
Our proof makes use of the well known notion of \emph{certificate complexity}.
\begin{definition}
An assignment $C: S\rightarrow\{0,1\}$ of values to some subset
$S\subseteq[n]$ of the $n$ variables is \emph{consistent} with
$x\in\01^n$ if $x_i=C(i)$ for all $i\in S$.                                            
For $b\in\01$, a \emph{$b$-certificate} for $f$ is an assignment $C$
such that $f(x)=b$ whenever $x$ is consistent with $C$.
The \emph{size} of $C$ is $|S|$, the cardinality of $S$.
The \emph{certificate complexity $C_x(f)$ of $f$ on $x$} is the size of
a smallest $f(x)$-certificate that is consistent with $x$.
The \emph{certificate complexity} of $f$ is $C(f)=\max_x C_x(f)$.
\end{definition}                                                                       %
\begin{lemma}
Let $p$ be an $\epsilon$-approximating polynomial for
$f:\01^n\rightarrow\01$, and $c=C(f)$ be the certificate complexity of $f$.
If $x\in\01^n$ and $z\in[0,1]^n$ satisfy $|x_i-z_i|\leq 1/(10c)$
for all $i\in[n]$, then $|p(z)-f(x)|\leq 6\epsilon/5+1/10$.
\end{lemma}
\begin{proof}
  Consider a certificate $C$ for $x$ of size $c$.  We will use $x^C$
  and $x^{\overline{C}}$ to denote the parts of $x$ corresponding to
  $C$ and to its complement, respectively, and write
  $x=x^Cx^{\overline{C}}$.  If $y\in\01^n$ is chosen according to the
  $z$-distribution ($y_i=1$ with probability $z_i$), then
  \begin{equation*}
    p(z)=\expect_y[p(y)]=
    \sum_{y^C y^{\overline C}} \Pr[ y^C ] \Pr [ y^{\overline C} ] 
    p(y^Cy^{\overline{C}})
    =
    \sum_{y^{\overline{C}}}\Pr[y^{\overline{C}}]\cdot
    \expect_{y^C}[p(y^Cy^{\overline{C}})].
  \end{equation*}
  Now consider the expectation
  $\expect_{y^C}[p(y^Cy^{\overline{C}})]$, where
  $y^{\overline{C}}\in\01^{n-c}$ is fixed, while the $y^C$-bits are
  still chosen according to the $z$-distribution.  Consider the
  $c$-variate polynomial obtained from $p$ by fixing the bits in
  $y^{\overline{C}}$. Since the ``error'' in the $z^C$-variables is at
  most $1/10c$, we have $\Pr[y^C=x^C]\geq (1-1/(10c))^c\geq 9/10$.
If $y^C\neq x^C$, then the difference between $p(y^Cy^{\overline{C}})$ 
and $p(x^Cy^{\overline{C}})$ is at most $1+2\epsilon$, so 
  $$
    |\expect_{y^C}[p(y^Cy^{\overline{C}})]-p(x^Cy^{\overline{C}})|
    \leq (1+2\epsilon)/10.
  $$
  But $f(x^Cy^{\overline{C}})=f(x)$, because the input
  $x^Cy^{\overline{C}}$ is consistent with the same certificate as $x$.  Hence
  \begin{eqnarray*}
   |\expect_{y^C}[p(y^Cy^{\overline{C}})]-f(x)| & \leq & |\expect_{y^C}[p(y^Cy^{\overline{C}})]-p(x^Cy^{\overline{C}})|+
   |p(x^Cy^{\overline{C}})-f(x)|\\
   & \leq & (1+2\epsilon)/10+\epsilon=1/10+6\epsilon/5,  
  \end{eqnarray*}
  and also $|p(z)-f(x)|\leq 6\epsilon/5+1/10$.
\qed
\end{proof}
This lemma implies that we can make a non-robust approximating
polynomial robust at the cost of a factor of $\bigO(\log C(f))$ in the
degree: replace each variable by an $\bigO(\log C(f))$-degree
amplification polynomial as used in the proof of Lemma~\ref{lem:typeoneboost}.  
Since it is known that $C(f)$ and $\approxdeg(f)$ are polynomially related
($C(f)=\bigO(\approxdeg(f)^4)$, see~\cite{buhrman&wolf:dectreesurvey}), we obtain:
\begin{theorem}
  $\robustdeg_{1,2}(f)=\bigO(\approxdeg(f)\cdot\log\approxdeg(f))$.
\end{theorem}

\section{Open Problems}
\label{sec:open-problems}

%
We mention some open problems. First, in contrast to
the classical case (\prbl{Parity}) we do not know of any function where
making a quantum algorithm robust costs more than a constant factor. 
Such a constant overhead suffices in the case of symmetric functions 
and functions whose approximate degree is $\Omega(n)$.
It is conceivable that quantum algorithms (and polynomials)
can \emph{always} be made robust at a constant factor overhead.
Proving or disproving this would be very interesting.

Second, we are not aware of a direct ``closed form'' or other natural way to
describe a robust degree-$n$ polynomial for the parity of $n$ bits,
but can only infer its existence from the existence of a robust quantum algorithm.
Given the simplicity of the non-robust representing polynomial for
\prbl{Parity}, one would hope for a simple closed form for robust polynomials
for \prbl{Parity} as well.

Finally, we have chosen our model of a noisy query such that we can
coherently make a query and reverse it. 
It is not clear to what extent non-robust quantum algorithms 
can be made resilient against decohering queries, since the usual 
transformations to achieve fault-tolerant quantum computation 
do not immediately apply to the query gate, which acts
on a non-constant number of quantum bits simultaneously.

\paragraph{Acknowledgments.}
We thank Peter H\o yer for inspiring initial discussions that 
led to our main result, and Michele Mosca for sending us a version of
\cite{ipy:biasedoracle}.
Oded Regev pointed out that when recovering all input bits, 
the quantum-search subroutine does not need to be robust.
Thanks to the anonymous TOCS referees for many helpful comments.

\bibliographystyle{plain}

\end{document}